# Ferroelectricity and ferromagnetism in EuTiO$_3$ nanowires

Anna N. Morozovska [*,1,2], Maya D. Glinchuk[1], Rakesh K. Behera[3], Basyl Y. Zaylichniy[1], Chaitanya S. Deo[3] and Eugene A. Eliseev[1,†]

[1]Institute for Problems of Materials Science, NAS of Ukraine,
Krjijanovskogo 3, 03142 Kiev, Ukraine

[2]V. Lashkarev Institute of Semiconductor Physics, NAS of Ukraine,
41, pr. Nauki, 03028 Kiev, Ukraine

[3] Nuclear and Radiological Engineering Program, George W. Woodruff School of Mechanical Engineering, Georgia Institute of Technology,
Atlanta, GA 30332, USA

**Abstract**

We predicted the ferroelectric-ferromagnetic multiferroic properties of EuTiO$_3$ nanowires and generated the phase diagrams in coordinates of temperature and wire radii. The calculations were performed within the Landau-Ginzburg-Devonshire theory with phenomenological parameters extracted from tabulated experimental data and first principles calculations. Since bulk EuTiO$_3$ is antiferromagnetic at temperatures lower than 5.5 K and paraelectric at all temperatures, our goal was to investigate the possibility of inducing the ferroelectric and ferromagnetic properties of EuTiO$_3$ by reducing the bulk to nanosystems.

Our results indicate that ferroelectric spontaneous polarization of $\sim 0.1$-$0.5 C/m^2$ is induced in EuTiO$_3$ nanowires due to the intrinsic surface stress, which is inversely proportional to the nanowire radius. The spontaneous polarization exists at temperatures lower than 300 K, for the wire radius less than 1 nm and typical surface stress coefficients $\sim 15$ N/m. Due to the strong biquadratic magnetoelectric coupling, the spontaneous polarization in turn induces the ferromagnetic phase at temperatures lower than 30 K for 2 nm nanowire, and at temperatures lower than 10 K for 4 nm nanowire in EuTiO$_3$. Thus we predicted that the EuTiO$_3$ nanowires can be the new ferroelectric-ferromagnetic multiferroic.

---
[*] Corresponding author1, morozo@i.com.ua;

[†] Corresponding author2 eliseev@i.com.ua

# I. Introduction

***1.1. Brief history of the question.*** Bulk quantum paraelectric EuTiO$_3$ is antiferromagnetic at temperatures lower than 5.5 K and paraelectric at all other temperatures [1, 2]. Oxide materials with large magnetoelectric (ME) coupling are very important for magnetoelectric applications based on the magnetic field control of the material dielectric permittivity [1, 2]. Although multiferroicity is a rare event, the search for new multiferroic is very important because of various applications.

Using *ab initio* calculations, Fennie and Rabe [3] predicted theoretically that (001) EuTiO$_3$ thin films subjected to the compressive epitaxial strain become ferromagnetic and ferroelectric simultaneously for strains exceeding 1.2%. Recently Lee *et al.* [4]] demonstrated experimentally that an epitaxial strain indeed turns EuTiO$_3$ into multiferroic. In particular, they demonstrated that EuTiO$_3$ thin film epitaxially grown on DyScO$_3$ substrate (corresponding to a tensile misfit strain of more that 1%) becomes ferromagnetic at temperatures lower than 4.24 K and ferroelectric at temperatures lower than 250 K. Lee *et al.* explained the appearance of the ferromagnetism in EuTiO$_3$ thin film by the strong spin–lattice biquadratic ME coupling.

***1.2. Motivation of our study.*** While the idea to induce the new multiferroic properties by elastic strain seems very attractive, its realization could meet some difficulties in epitaxial films, since relatively high misfit strains (~1-3 %) between the film and substrate relaxes because of e.g. the appearance of misfit dislocations [5, 6]. It is extremely difficult to synthesize a strongly strained epitaxial film without rather special, complex and thus high-cost deposition processes. It has been demonstrated that strain relaxation to the values lower than 1% eliminates ferroelectric-ferromagnetic phase appearance in EuTiO$_3$ thin film [3].

However, in nanowires and nanorods (**Fig.1a**), the intrinsic surface stress exists spontaneously due to the surface curvature and typically does not relax in nanowires and nanorods. Surface stress is inversely proportional to the wire radius and directly proportional to the surface stress tensor (similar to Laplace surface tension). The intrinsic surface stress should depend both on the growth conditions and the surface termination morphology [7, 8]. Although surface tension appears even for the case of non-reconstructed geometrical surfaces due to the surface curvature [8], surface reconstruction should affect the surface tension value or even be responsible for the appearance of surface stresses [9, 10].

It is shown that the intrinsic surface stress can induce ferroelectricity, ferromagnetism and increase corresponding phase transition temperatures in conventional and quantum paraelectric nanorods and nanowires [11, 12, 13, 14, 15]. These facts motivated us to explore the possibility of inducing simultaneous ferroelectricity and ferromagnetism in EuTiO$_3$ nanowires due to the intrinsic surface stress. In this article we performed analytical calculations of EuTiO$_3$ nanowires by exploring the ferroelectic, magnetic properties and phase diagrams within conventional Landau-Ginzburg-Devonshire (LGD) theory with phenomenological parameters extracted from first principles calculations and tabulated experimental data.

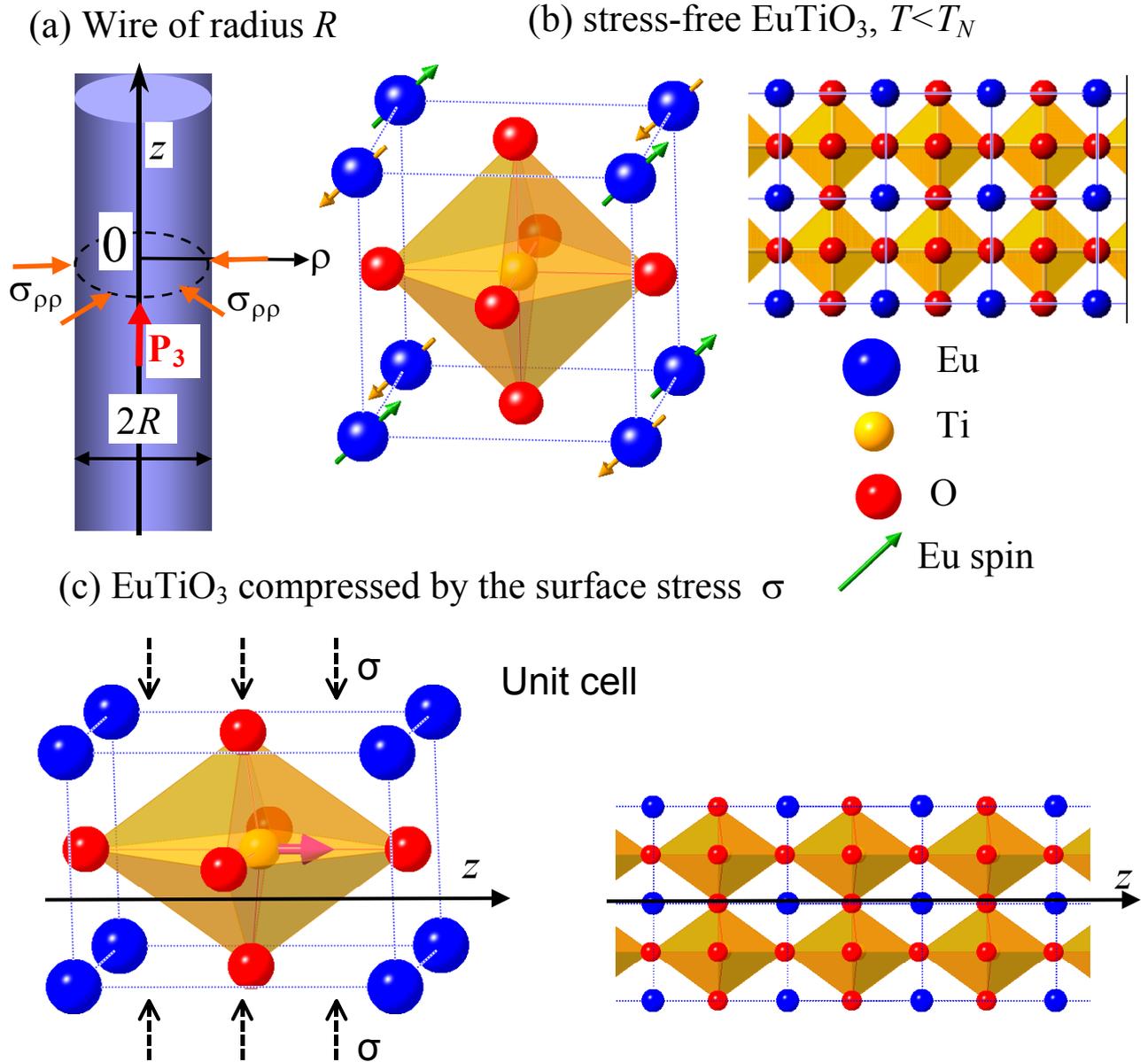

**Fig. 1**. (Color online) (a) Schematics of a high aspect ratio nanowire, where $P_3$ is the polarization along the z-direction and $\rho$ is the polar radius, $\sigma_{\rho\rho}$ is the intrinsic surface stress. (b) The stress-free EuTiO$_3$ unit cell in a bulk material in the antiferromagnetic phase. (c) EuTiO$_3$ unit cells subjected to the surface stress (Laplace tension). Solid pink arrow indicates the polarization direction.

***1.3. Background of the used method.*** The application of the continuum media LGD theory to the description of nanowire polar and magnetic properties requires justification due to the small size of the object of study. The continuum media theory was successfully used for the analysis of elastic properties of metallic, semiconductor, dielectric or polymeric nanowires and nanotubes [16, 17, 18, 19] and the piezoelectric response [20]. For nanosized ferroics, the applicability of the continuous media phenomenological theory is corroborated by the fact that the critical sizes (~2-10 lattice constants) of the appearance of long-range order calculated from atomistic [21] and phenomenological theories [13, 14, 15] are in **good agreement** with each other [22, 23] as well as with experimental results [24]. The long-range order appears for sizes larger than critical ones. Once the long-range order is established, it is possible to apply the mean field LGD theory [25]. Thus the agreement between the magnitudes of the critical sizes calculated from LGD and atomistic theories are extremely important

**II. Basic Formalism**

LGD free energy $F$ depends on the polarization vector **P**, magnetic sublattices (a,b) magnetization vector $\mathbf{M} = (\mathbf{M}_a + \mathbf{M}_b)/2$ and antimagnetization vector $\mathbf{L} = (\mathbf{L}_a - \mathbf{L}_b)/2$ as:

$$F = F_P + F_M + F_{ME} \tag{1a}$$

where $F_P$ is polarization-dependent, $F_M$ is magnetization-dependent and

$$F_{ME} = \frac{P^2}{2}\left(\eta_{FM} M^2 + \eta_{AFM} L^2\right) \tag{1b}$$

is the biquadratic magnetoelectric (**ME**) coupling energy. The **ME** coupling coefficients $\eta_{FM} < 0$ for ferromagnetics and $\eta_{AFM} > 0$ for antiferromagnetics [4]. Following Lee *et al.* [4] we can regard that $\eta_{AFM} \approx -\eta_{FM} > 0$ for numerical calculations, as anticipated for equivalent magnetic Eu ions with antiparallel spin ordering in a bulk EuTiO$_3$ (see **Fig.1b**).

Considering very long cylindrical EuTiO$_3$ nanowires with polarization and external electric field (if any) directed along the cylinder axes $z$ one could neglect the depolarization field if the wire length $h$ is much higher than its radius $R$, namely $h/R >> 10^3$ [11]. Thus we considered very high aspect ratio wires, for which the polarization vector $\mathbf{P} = (0,0,P_3(\rho))$ appears in the z-direction along the wire axis, and $P_3(\rho)$ depends only on the polar coordinate $\rho = \sqrt{x^2 + y^2}$ due to the radial symmetry of the problem (as shown in **Fig.1a**). The single-component polarization $P_3$ and elastic strains $u_{ij}$ dependent part of the free energy is

$$F_P = \left( \begin{array}{c} \int_V d^3r \left( \frac{\alpha_P(T)}{2}P_3^2 + \frac{\beta_P}{4}P_3^4 + \frac{\gamma_P}{6}P_3^6 + \frac{g}{2}(\nabla P_3)^2 - P_3 E_3^e + c_{ijkl} u_{ij}\left(\frac{u_{kl}}{2} + Q_{kl33}P_3^2\right) \right) \\ + \int_S d^2r \left( \frac{\alpha_P^S}{2}P_3^2 + \mu_{ij}^S u_{ij} + d_{ij3}^{Se} u_{ij} P_3 \right) \end{array} \right) \tag{1c}$$

Integration in Eq.(1c) is performed over the system volume $V$ and surface $S$. Following are the description of the variables in Eq.(1c): $E = (0,0, E_3^e)$ is the external electric field. The coefficient $\gamma_P > 0$ and the gradient term g > 0. In accordance with Barrett law, that has to be applied for quantum paraelectric EuTiO$_3$, the coefficient $\alpha_P(T) = \alpha_T\left(\coth(T_q/(2T))T_q/2 - T_c\right)$, where $T$ is the absolute temperature, $T_c$ is the Curie temperature (negative for EuTiO$_3$ [1]) as it has to be for quantum paraelectric, $T_q$ is the quantum vibration temperature, which is positive. $c_{ijkl}$ is the stiffness tensor and $Q_{kl33}$ is the electrostriction tensor component.

For correct phenomenological description of the spatially confined system the surface energy should be considered, the contribution of which increases with the decrease in the wire radii. The surface energy coefficient $\alpha_P^S \geq 0$ is isotropic and weakly temperature dependent. Thus, higher terms can be neglected in the surface energy expansion. $\mu_{ij}^S$ is the intrinsic surface stress tensor that exists under the curved surface of solid body and determines the excess pressure on the surface [8]. We will use the isotropic approximation $\mu_{ij}^S = \mu \delta_{ij}$, where the scalar coefficient $\mu$ varies in the typical range of 3–30 N/m [26, 27, 28, 29]. This surface stress coefficient can be manipulated in some range by the choice of the ambient of the nanowire (template material, composite glassy matrix, glue, gel or gas), and depend on surface reconstruction. $d_{ij3}^{Se}$ is the tensor of the surface piezoelectric effect originated from the disappearance of inversion symmetry in the vicinity of the surface [30, 31].

The magnetization-dependent part of the free energy is

$$F_M = \int_V d^3r \left( \begin{array}{c} \dfrac{\alpha_M(T)}{2}M^2 + \dfrac{\alpha_L(T)}{2}L^2 + \dfrac{\beta_M}{4}M^4 + \dfrac{\beta_L}{4}L^4 + \dfrac{\lambda}{2}L^2M^2 ... \\ -\mu_0 H_i M_i + c_{ijmn}u_{ij}\left(Z_{mnkl}M_k M_l + \widetilde{Z}_{mnkl}L_k L_l\right) \end{array} \right) \qquad (1d)$$
$$+ \int_S d^2r \left( \dfrac{\alpha_M^S}{2}M^2 + d_{ijk}^{Sm}u_{ij}M_k \right)$$

Coefficients $\alpha_M(T) = \alpha_C(T - T_C)$ and $\alpha_L(T) = \alpha_N(T - T_N)$, $\mu_0 = 4\pi \cdot 10^{-7}$ Henri/m is the universal magnetic constant, $M^2 = M_1^2 + M_2^2 + M_3^2$ is the magnetization, and $L^2 = L_1^2 + L_2^2 + L_3^2$ is the antimagnetization values, **H** is an external magnetic field (if any), $Z_{mnkl}$ and $\widetilde{Z}_{mnkl}$ stands for magnetostriction and antimagnetostriction tensors respectively. For equivalent permutated magnetic Eu ions with antiparallel spin ordering it can be assumed that $\alpha_C \sim \alpha_N$. Note, that $\alpha_M(T) = \alpha_C(T - T_C)$ determines the experimentally observed magnetic susceptibility in paramagnetic phase [1, 2, 4]. The positive coupling term $\dfrac{\lambda}{2}L^2M^2$ prevents the appearance of ferromagnetic (as well as ferrimagnetic) phase at temperatures $T < T_C$ under the condition $\sqrt{\beta_M \beta_L} < \lambda$ (see Suppl.Materials, Appendix B). Hereinafter, this condition is assumed to be valid for the current study. The last integral in Eq.(1d) is the magnetic surface energy including the surface piezomagnetic effect (via the corresponding tensor $d_{ijk}^{Sm}$) that can exist at least at low temperatures [32]. All EuTiO$_3$ parameters involved in Eqs.(1) are listed in **Table 1.** The parameters were extracted from the fitting to experimental data (EXP) and first principles calculations (DFT). Note that no ferromagnetic phase were reported [1] for EuTiO$_3$, thus we suppose the condition $\sqrt{\beta_M \beta_L} < \lambda$ to be valid.

**Table 1**. LGD free energy expansion coefficients for quantum paraelectric bulk material EuTiO$_3$

| Parameter | Unit | Quantum paraelectric EuTiO$_3$ | Notes and Refs |
|---|---|---|---|
| * Background permittivity $\varepsilon_b$ | Dimensionless | 181 | [1, [33] EXP |
| LGD-coefficient $\alpha_T$ | m/(F K) | $4.83 \times 10^6$ | [1, 34, 35] EXP |
| FE Curie temperature $T_c$ | K | $-25$ | [1, 33, 34] EXP |
| Quantum vibration temperature $T_q$ | K | 162 | [1, 33, 34] EXP |
| LGD-gradient coefficient $g$ | V·m$^3$/C | $\sim 10^{-10}$ | [36] EXP |
| LGD-coefficient $\beta_P$ | m$^5$/(C$^2$F) | $5 \times 10^9$ | fitting results of [3, 4] DFT |
| LGD-coefficients $\gamma_P$ | m$^9$/(C$^4$F) | $\ll 10^{11}$ | Not enough data to determine exactly |
| Electrostriction coefficients $Q_{ijkl}$ (Voigt notation) | m$^4$/C$^2$ | $Q_{11} = 0.13$ $Q_{12} = -0.035$ | fitting results of [3, 4] DFT |
| † Elastic stiffness $c_{ij}$ | N/m$^2$ | $c_{11} = (3.27 \pm 0.1) \times 10^{11}$ $c_{12} = (1.04 \pm 0.03) \times 10^{11}$ | DFT [this paper] |
| # Elastic compliances $s_{ij}$ | m$^2$/N | $s_{11} = (3.62 \pm 0.1) \times 10^{-12}$ $s_{12} = -(0.87 \pm 0.2) \times 10^{-12}$ | DFT [this paper] |
| LGD-coefficient $\alpha_C \approx \alpha_N$ | Henri/(m·K) | $2\pi \cdot 10^{-6}$ | [4, 1] EXP |
| LGD-coefficient $\beta_M$ | J m/A$^4$ | $0.8 \times 10^{-16}$ | fitting results of [1] EXP |
| LGD-coefficient $\beta_L$ | J m/A$^4$ | $1.33 \times 10^{-16}$ | fitting results of [1] EXP |
| LGD-coefficient $\lambda$ | J m/A$^4$ | $1.0 \times 10^{-16}$ | fitting results of [1] EXP |
| Magnetostriction coefficients $Z_{ij}$ (Voigt notation) | m$^2$/A$^2$ | $Z_{12} = -(5.25 \pm 0.75) \times 10^{-16}$ $Z_{11} = (8.75 \pm 1.75) \times 10^{-16}$ | fitting results of [3, 4] DFT |
| Magnetostriction coefficients $\widetilde{Z}_{ij}$ (Voigt notation) | m$^2$/A$^2$ | $\widetilde{Z}_{12} = -(5.25 \pm 0.25) \times 10^{-16}$ $\widetilde{Z}_{11} = (7.75 \pm 0.75) \times 10^{-16}$ | fitting results of [3, 4] DFT |
| AFM Neel temperature $T_N$ | K | 5.5 | [2] EXP |
| FM Curie temperature $T_C$ | K | $3.5 \pm 0.3$ | [1,2] EXP |
| Biquadratic ME coupling coefficient $\eta_{AFM} = -\eta_{FM}$ | J m$^3$/(C$^2$ A$^2$) | $0.16 \times 10^{-3}$ | fitting results of [1] EXP |

* Full permittivity $\varepsilon = \left( \varepsilon_b + \dfrac{1}{\alpha \varepsilon_0} \right)$, soft mode related permittivity $\varepsilon_{QP} = (\alpha \varepsilon_0)^{-1}$

† Elastic stiffness values was calculated in Appendix D in the Suppl. Materials from the bulk modules $(173 \pm 10)$GPa extracted from Ranjan *et al.* [37] DFT data.

# Elastic compliance matrix is calculated from the stiffness matrix as inverse matrix

Note, that the surface piezoeffects coupled with the intrinsic surface stress can favor the long-range order appearance in the nanosized system [32]. Since no reliable data is known about the surface piezoelectric $d_{ij3}^{Se}$ and piezomagnetic $d_{ij3}^{Sm}$ coefficients and surface energy coefficients $\alpha_M^S$ and $\alpha_P^S$, we did not include these into the numerical calculations. It was shown [11-13, 38], that for the case $\alpha_{M,P}^S = 0$, corresponding polarization, magnetization and strain spatial distributions become quasi-homogeneous inside the wire cross-section, i.e. almost independent of the polar coordinate ρ, but their average values are dependent on the wire radius $R$. Thus we have considered the case $\alpha_{M,P}^S = 0$, which allowed us to list and analyze only the average values of the properties and do not concentrate on their spatial distributions.

The nonzero components of the strain tensor inside a cylindrical wire of radius $R$ subjected to the intrinsic surface stress, electrostriction and magnetostriction effects have the following form:

$$u_{11} = -(s_{11}+s_{12})\frac{\mu}{R} + Q_{12}P_3^2 + Z_{12}(M_3^2 + M_2^2) + Z_{11}M_1^2 + \tilde{Z}_{11}L_1^2 + \tilde{Z}_{12}(L_3^2 + L_2^2), \quad (2a)$$

$$u_{22} = -(s_{11}+s_{12})\frac{\mu}{R} + Q_{12}P_3^2 + Z_{12}(M_3^2 + M_1^2) + Z_{11}M_2^2 + \tilde{Z}_{11}L_2^2 + \tilde{Z}_{12}(L_3^2 + L_1^2), \quad (2b)$$

$$u_{33} = -s_{12}\frac{\mu}{R} + Q_{11}P_3^2 + Z_{11}M_3^2 + Z_{12}(M_1^2 + M_2^2) + \tilde{Z}_{11}L_3^2 + \tilde{Z}_{12}(L_1^2 + L_2^2). \quad (2c)$$

Here $s_{ij}$'s are the elastic compliances inherent to the bulk material. It is important to mention that the elastic compliances for nanowire and thin films may be different from those of the bulk material [39, 40, 41], but LGD deals with bulk coefficients and obtain their renormalization self-consistently.

Elastic strains calculated from Eqs.(2) are shown in **Fig.2**. For wire radii $R$ >2.5 nm the absolute value of compressive strains $u_{11,22}$ becomes less than 1.2% and should lead to the disappearance of ferroelectricity, which is in agreement with Ref. [3].

The difference between the axial tensile $u_{33}$ and compressive $u_{\rho\rho}$ strains radial dependencies are rather small, but visible (compare top and bottom curves in Fig.2). Namely, the small impact of the electrostriction coupling is seen for the tensile strain $u_{33}$ only (breaks on blue and red curves appeared when the spontaneous polarization appears at the critical radius), while the peculiarities are almost invisible for $u_{\rho\rho}$ due to the fact that the electrostriction coefficient $|Q_{12}|$ is several times smaller than $Q_{11}$. The magnetostriction contribution to both lateral and longitudinal strains is very small in comparison with the strain caused by the surface tension and electrostriction effect [13]. Thus the approximation $u_{11} \approx u_{22} \approx u_{\rho\rho} \approx -(s_{11}+s_{12})(\mu/R)+Q_{12}P_3^2$, $u_{33} \approx -s_{12}(\mu/R)+Q_{11}P_3^2$ may be used for estimations (see also Refs.[11, 42]). The effect $u_{\rho\rho} \neq u_{33}$, represented by Eqs.(2), leads to the unit cell tetragonality, namely to its compression in the radial direction (since $u_{\rho\rho} < 0$) and elongation along the wire axes (since $u_{33} > 0$) [42]. The strain radius dependence is well fitted by the law $u_{ij} \sim \mu/R$, proving that the strains are primarily originated from the intrinsic surface stress.

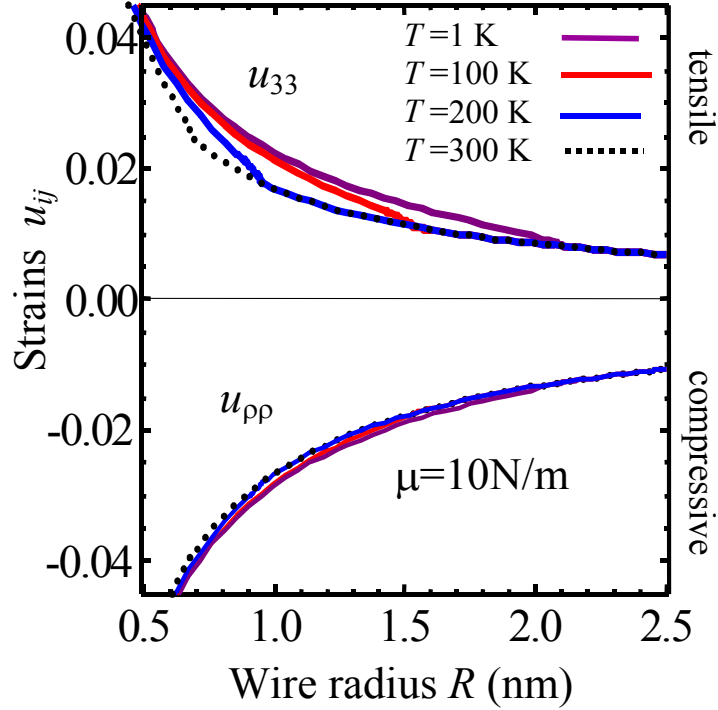

**Fig. 2**. (Color online) Strains $u_{ij}$ vs. radius $R$ calculated for the surface stress $\sigma_{\rho\rho} = -\mu/R$ with coefficient $\mu$=10 N/m and different temperatures $T$ = 1, 100, 200, 300 K. The intrinsic surface stress is compressive for thermodynamically stable surfaces and the component $\sigma_{\rho\rho} = -\mu/R$ has negative sign as directed inside the wire cross-section (see **Fig.1a** and Refs.[11-13, 25]).

Using the expressions for strains and direct variational method [11-13], the spontaneous polarization $\overline{P}_3(T,R)$, magnetization $\overline{M}(T,R)$ and antiferromagnetic order parameter $\overline{L}(T,R)$ averaged over the wire radius $R$ was derived as:

$$\overline{P}_3(T,R) \approx \sqrt{-\frac{\alpha_R(T,R)}{\beta_P}}, \qquad M = L = 0, \qquad (3a)$$

(in ferroelectric-paramagnetic **FE+PM** phase)

$$\overline{P}_3(T,R) \approx \sqrt{-\frac{\alpha_R - \eta_{FM}(\alpha_{MR}/\beta_M)}{\beta_P - (\eta_{FM}^2/\beta_M)}}, \quad \overline{M}(T,R) \approx \sqrt{-\frac{\alpha_{MR} - \eta_{FM}(\alpha_R/\beta_P)}{\beta_M - (\eta_{FM}^2/\beta_P)}}, \quad L=0, \qquad (3b)$$

(in ferroelectric-ferromagnetic **FE+FM** phase)

$$\overline{P}_3(T,R) \approx \sqrt{-\frac{\alpha_R - \eta_{AFM}(\alpha_{LR}/\beta_L)}{\beta_P - (\eta_{AFM}^2/\beta_L)}}, \quad \overline{L}(T,R) \approx \sqrt{-\frac{\alpha_{LR} - \eta_{AFM}(\alpha_R/\beta_P)}{\beta_L - (\eta_{AFM}^2/\beta_P)}}, \quad M=0, \qquad (3c)$$

(in ferroelectric-antiferromagnetic magnetic **FE+AFM** phase)

In Eqs.(3) we considered $\gamma_P = 0$. The coefficients $\alpha_R(T,R)$, $\alpha_{MR}(T,R)$ and $\alpha_{LR}(T,R)$ are given by expressions

$$\alpha_R(T,R) \approx \alpha_T \left( \frac{T_q}{2} \coth\left(\frac{T_q}{2T}\right) - T_c \right) + Q_{12}\frac{4\mu}{R}, \qquad (4a)$$

$$\alpha_{MR}(T,R) \approx \alpha_C(T-T_C) + W\frac{4\mu}{R}, \quad \alpha_{LR}(T,R) \approx \alpha_N(T-T_N) + \widetilde{W}\frac{4\mu}{R}. \qquad (4b)$$

It is seen from the **Table 1** that the inequalities $\widetilde{Z}_{12} < 0$, $Z_{12} < 0$, $\widetilde{Z}_{12} + \widetilde{Z}_{11} > 0$, $Z_{12} + Z_{11} < 0$ are valid. So, when the spontaneous (anti)magnetization is directed along the z-axes, the parameters in Eqs.(4b) are $\widetilde{W} = +\widetilde{Z}_{12}$, $W = +Z_{12}$. Similarly, when the spontaneous (anti)magnetization is lying in the {x,y} plane, the parameters in Eqs.(4b) are $\widetilde{W} = -(\widetilde{Z}_{12} + \widetilde{Z}_{11})/2$, $W = -(Z_{12} + Z_{11})/2$.

Using the condition $\overline{P}_3(T,R) = 0 \leftrightarrow \alpha_R(T,R) = 0$ (see Eq.(3a)) and expression for the coefficient $\alpha_R(T,R)$ given by Eq.(4a), the transition temperature from paraelectric-paramagnetic (**PM+PE**) into the ferroelectric-paramagnetic (**FE+PM**) phase $T^{FE}(R)$ was derived as:

$$T^{FE}(R) \approx \frac{T_q}{2}\mathrm{arccoth}^{-1}\left(\frac{2}{T_q}\left(T_c - Q_{12}\frac{4\mu}{\alpha_T R}\right)\right), \qquad (5)$$

The form of Eq.(5) follows from the Barrett law; the positive term $-Q_{12}\frac{4\mu}{\alpha_T R}$ is the contribution of intrinsic surface stress $\sigma_{\rho\rho} = -\mu/R$. The second term in Eq.(5) increases $T^{FE}(R)$, since $-Q_{12} > 0$ (see **Table 1**).

Using the conditions $\overline{M}(T,R) = 0$ and $\overline{L}(T,R) = 0$ in Eqs.(3b,c) and expressions for the coefficients $\alpha_{MR}(T,R)$ and $\alpha_{LR}(T,R)$ given by Eqs.(4b), the self-consistent equations for the determination of the transition temperature from the paramagnetic (**PM**) in the ferromagnetic (**FM**) and antiferromagnetic (**AFM**) phases were derived as:

$$T^{FM} \approx T_C - W\frac{4\mu}{\alpha_C R} + \frac{\eta_{FM}}{\alpha_C}\frac{\alpha_R(T^{FM},R)}{\beta_P}, \qquad (6a)$$

$$T^{AFM} \approx T_N - \widetilde{W}\frac{4\mu}{\alpha_N R} - \frac{\eta_{AFM}}{\alpha_N}\frac{\alpha_R(T^{AFM},R)}{\beta_P}. \qquad (6b)$$

Here $\overline{P}_3^2(T,R) \approx -\frac{\alpha_R}{\beta_P}$ in accordance with Eq.(4a). Note, that $T_N - T_C \approx 1.7\,\text{K}$ for bulk EuTiO$_3$ [1, 2]. Following *Lee et al.* [4], we assigned $\eta_{AFM} = -\eta_{FM} > 0$ and $\widetilde{Z}_{ij} \approx Z_{ij}$, $\alpha_C \approx \alpha_N$, so that the expression for the difference in transition temperatures is given as:

$$T^{FM}(R) - T^{AFM}(R) \approx T_C - T_N + \frac{\eta_{AFM}}{\alpha_N}\left(\frac{\alpha_R(T^{FM},R)}{\beta_P} + \frac{\alpha_R(T^{AFM},R)}{\beta_P}\right). \qquad (7)$$

The last term in Eq.(7) becomes positive because $\eta_{AFM} > 0$ [4] and $\alpha_N$ is always positive. So the condition $T^{FM}(R) > T^{AFM}(R)$ can be reached for EuTiO$_3$ nanowires, as observed experimentally for EuTiO$_3$ thin strained films [4].

Similarly, the boundary between the **FM+FE** phase and **FM+PE** phase is determined from the condition $\overline{P}_3(T,R) = 0$ in Eq.(3b) as $\alpha_R(T,R) - \eta_{FM}(\alpha_{MR}(T,R)/\beta_M) = 0$. The boundary between the **AFM+FE** phase and **AFM+PE** phase is determined from the condition $\overline{P}_3(T,R) = 0$ in Eq.(3c) as $\alpha_R(T,R) - \eta_{AFM}(\alpha_{LR}(T,R)/\beta_L) = 0$.

Following the expressions in Eqs.(3-6) it is obvious that by changing the wire radius one can control the phase diagram (e.g. FE, AFM and FM phase transition temperatures), and corresponding spontaneous polarization and magnetization values in the EuTiO$_3$ nanowires.

The possibility of tuning spontaneous polarization is illustrated in **Figs.3a,b**. As it is seen from **Fig.3a,** spontaneous polarization $\overline{P}_3(T,R)$ increases with the decrease in wire radius for a fixed surface tension coefficient $\mu$. For $\mu$=10 N/m, EuTiO$_3$ nanowire of radius ~2 lattice constants (1 *l.c.*≈0.4 nm) is predicted to be ferroelectric at temperatures lower than 300 K. The wire of radius ~4 lattice constants is predicted to be ferroelectric at temperatures lower than 100 K (**Fig. 3(b)**). For $\mu$=10 N/m and wire radius of 2-4 lattice constants (~0.8 – 1.6 nm), the spontaneous polarization of the nanowire reaches the values of ~0.5-0.1 C/m$^2$ at temperatures lower than 100 K. Note, that LGD phenomenology predicts higher enhancement of $\overline{P}_3(T,R)$ values for nanowires with radius $R = 1$ lattice constant, but LGD continuum approach is not quantitatively correct for such small size.

The cusp or peculiarity on polarization temperature dependence (marked by filled circles in **Figs.3b**) indicates the phase transition related to the appearance of magnetization *M* or antimagnetization *L* (as shown by abrupt changes on the curves in **Figs.3d**). For $\mu$=10 N/m the cusp appeared at temperatures $T \approx 27, 12, 10, 6$ K for the wire radius $R \approx 1, 2, 3, 6$ lattice constants respectively. The peculiarity is the direct manifestation of the ME coupling.

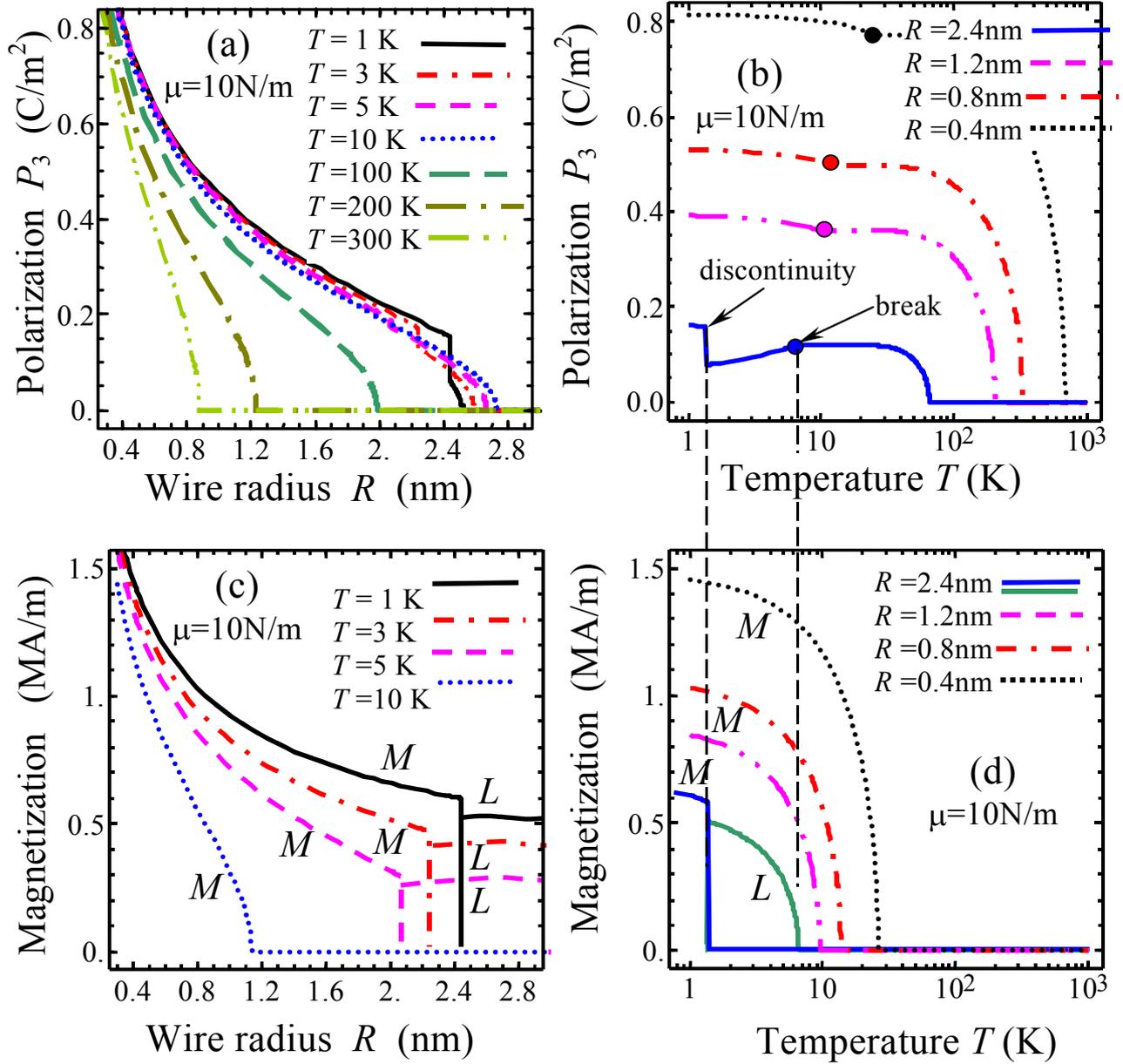

**Fig. 3**. (Color online) Variation of (a) spontaneous polarization with wire radius calculated at different temperatures (*T* = 1, 3, 5, 10, 100, 200, 300 K), (b) spontaneous polarization with temperature calculated at different wire radii (*R* ≈ 1, 2, 3, 6 lattice constants, 1 *l.c.* ≈ 0.4 nm), (c) magnetization (letter "*M*") and antimagnetization (letter "*L*") with wire radius calculated at different temperatures (*T* = 1, 3, 5, 10 K), and (d) M and L with temperature calculated at different wire radii (*R* ≈ 1, 2, 3, 6 lattice constants). The cusp or peculiarity on polarization temperature and radius dependence is marked by filled circles in **Fig.3b**. Surface tension coefficient μ=10 N/m for all plots.

The possibility of tuning spontaneous magnetization is explored in **Figs.3c,d.** As it is seen from **Fig.3c** the spontaneous magnetization $M(T,R)$ increases with the decrease in wire radius. For μ=10 N/m, EuTiO$_3$ nanowire of radius ~2-6 lattice constants is ferromagnetic ($M \neq 0$) at low temperatures and antiferromagnetic ($L \neq 0$) at higher sizes. The spontaneous magnetization of the nanowire reaches the values more than 1 MA/m at low temperatures. Note, that the values more than 1 MA/m are excluded in practice by the net magnetic moment of Eu ($7g\mu_B/2$). Realistic magnetization curves should saturate prior to reaching the limiting value. Rigorously speaking, the limitation should be accounted in the LGD thermodynamic theory e.g. by adding magnetization expressions of higher than the fourth power in Eq.(1d) [43]. We omitted the higher magnetization powers in Eq.(1d) in order to obtain analytical expressions for magnetization and antimagnetization given by Eq.(3b,c), which nevertheless are rather rigorous in the vicinity of the magnetic phase transition boundaries.

Phase diagram of the EuTiO$_3$ 1nm-nanowire ($R \sim 2.5$ lattice constants) in coordinates of temperature $T$ − surface stress coefficient µ is illustrated in **Fig.4a** for the temperature range of 0 to 300 K. **Figure 4b** is a magnified in view of Fig.4a for temperatures lower than 30 K, which shows the multiferroic phase boundaries at lower temperatures. It is seen that the FE+PM, FM+FE and AFM+FE phases appear in the nanowires subjected to the intrinsic surface stress $\sigma_{\rho\rho} = -\mu/R$, in contrast to the bulk material with σ=0, which can attain PE+PM and AFM+PE phases only. The ferroelectric and ferromagnetic phase transition temperatures increase with the increase in the surface stress, which in turn is inversely proportional to the wire radius in the continuum theory.

The possibility of the transition temperatures tuning by changing the nanowire radius is explored in **Figs.4c-d** for two different surface tension coefficient µ=10 N/m and 30 N/m. For µ=10 N/m the surface stress and ME coupling induce the radius-dependent FE+FM phase in EuTiO$_3$ nanowires of radius less than 3 lattice constants at temperatures lower than 10 K. For µ=30 N/m the surface stress and ME coupling induce the radius-dependent FE+FM phase in EuTiO$_3$ nanowires of radius less than 3 lattice constants at temperatures lower than 30 K. For µ=30 N/m and radius less than 10 lattice constants (~4 nm), FE+FM phase appears at temperatures lower than 10 K. Thus, the higher is the coefficient µ, the wider is the region of the multiferroic FE+FM phase (compare plots c, d). The region of the multiferroic phase increases with the decrease in wire radius.

In Figs. 4b,c the boundary between PM+PE and PM+FE phase appeared to be vertical at temperatures lower than 30 K. This indicates that the boundary becomes virtually independent on temperature in this condition. This is because at temperatures essentially lower than $T_q$ (that is about 162 K for EuTiO$_3$) the coefficient $\alpha_R(T,R) \approx \alpha_T\left(\coth(T_q/(2T))T_q/2 - T_c\right) + Q_{12}(4\mu/R)$ becomes independent of temperature. The temperature independence of $\alpha_R(T,R)$ at $T \ll T_q$ in turn originated from the temperature "plateau" at $T \ll T_q$ for the coefficient $\alpha_P(T) = \alpha_T\left(\coth(T_q/(2T))T_q/2 - T_c\right)$, which is in accordance with the Barrett law.

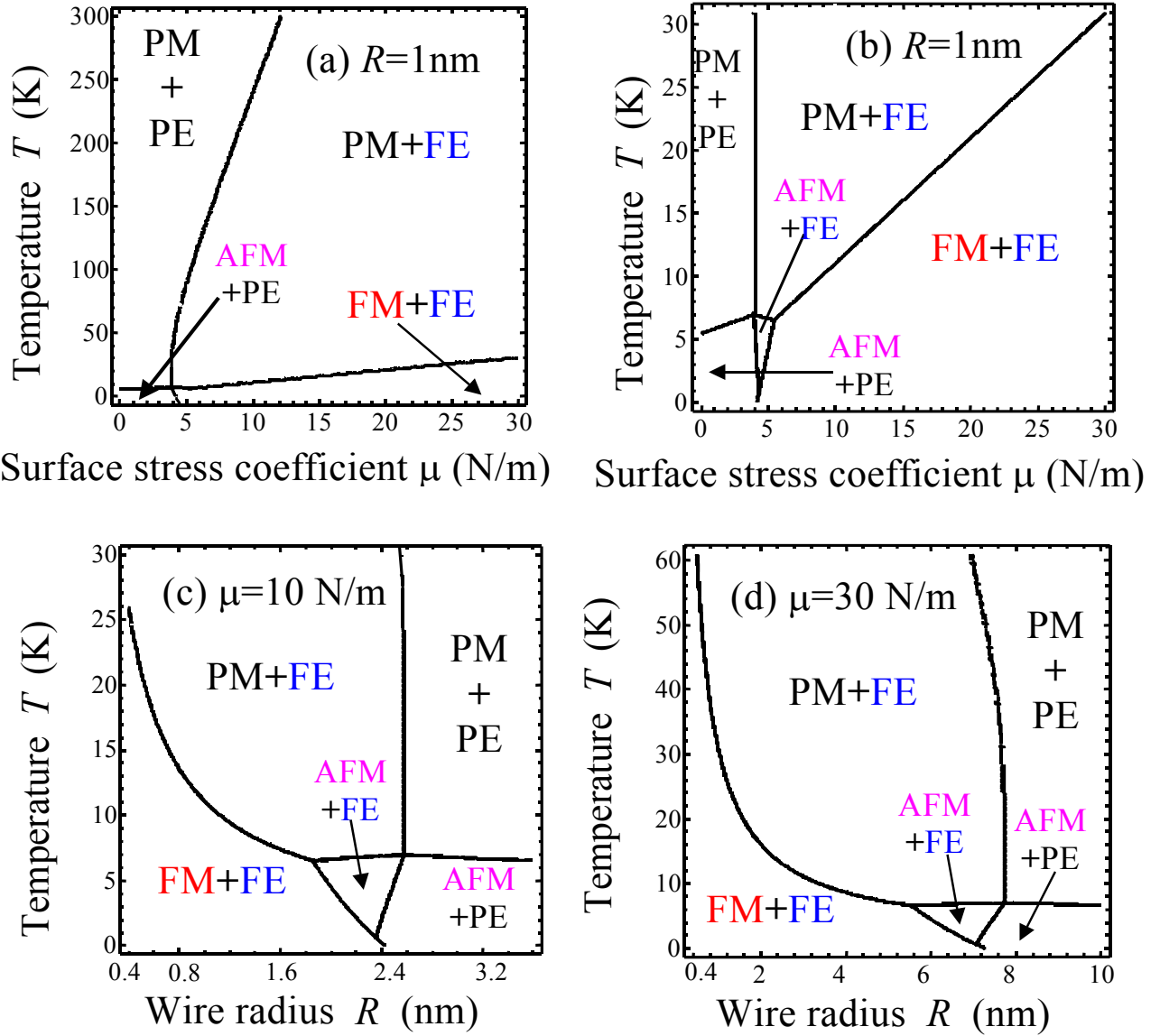

**Fig. 4**. (a,b) Phase diagram of EuTiO$_3$ nanowire in coordinates of temperature $T$ − surface stress coefficient μ for the wire radius $R$= 1 nm [(a) 0<$T$<300 K and (b) 0<$T$<30 K]. (c,d) Phase diagrams of the EuTiO$_3$ nanowires in coordinates temperature $T$- wire radius $R$ calculated for different surface stress coefficient μ= 10, and μ= 30 N/m. AFM − antiferromagnetic phase, FM − ferromagnetic phase, PM − paramagnetic phase, FE − ferroelectric phase, PE − paraelectric phase.

It is evident from our theoretical analysis that changing EuTiO$_3$ from bulk to nanowire it is possible to induce multiferroic properties. Manipulating the wire radii and surface stresses, the multiferroic nature of these nanowires can be tuned. Quantitatively, we demonstrated how the spontaneous polarization (induced by the intrinsic surface stress) in turn induces the ferromagnetic phase in EuTiO$_3$ nanowires with radii less than 1-10 lattice constants (~0.4-4 nm) due to the biquadratic ME coupling.

**Summary**

Using Landau-Ginzburg-Devonshire free energy with material parameters extracted from the first principle calculations and experimental data, we calculated the radii and temperature dependence of the paraelectric, ferroelectric, paramagnetic, ferromagnetic, antiferromagnetic and multiferroic phases in EuTiO$_3$ nanowires.

Ferroelectric spontaneous polarization of about ~(0.1–0.5)C/m$^2$ is induced by the intrinsic surface stress present under the curved surface of the nanowire. Since the stress is inversely proportional to the nanowire radius, the spontaneous polarization exists for the radii less than (2–5) nm, for acceptable values of the surface stress coefficient ~ (10–30) N/m and temperatures lower than 300 K.

Due to the strong biquadratic magnetoelectric coupling, the spontaneous polarization in turn induces the ferromagnetic phase in EuTiO$_3$ nanowires at temperatures lower than 30 K. Thus we predicted that the EuTiO$_3$ nanowires becomes ferroelectric-ferromagnetic, i.e. multiferroic, which can be important for their potential applications.

Regarding the applications, multiferroic ferromagnetic-ferroelectric EuTiO$_3$ nanowires potentially can win the competition with strained EuTiO$_3$ thin films, which becomes ferromagnetic at temperatures lower than 4.24 K, since we predicted ferromagnetism at temperatures lower than 30 K in the nanowires. In general, nanosized EuTiO$_3$ wires as new multiferroic is favourable in comparison with a bulk material, since bulk EuTiO$_3$ is antiferromagnetic at temperatures less than 5.5 K and paraelectric at all other temperatures, i.e. it has neither polarization nor magnetization.

We hope that our prediction can stimulate both synthesis of the EuTiO$_3$ nanowires (e.g. in the form of nanopowders or arrays embedded into the different matrices) and experimental studies of their ferroic properties as well as the first-principles calculations of the spontaneous dipole moment and magnetization induced by the intrinsic stress under the curved surface.

# Supplementary materials to
# "Ferroelectricity and ferromagnetism in EuTiO$_3$ nanowires"

In the main article, we have reported the multiferroic nature of EuTiO$_3$ nanowires based on wire radii, temperature, and surface characteristics (surface stress and surface stress coefficient) using Landau-Ginzburg-Devonshire (LGD) theory. All the values estimated from the LGD continuum approach are based on the bulk EuTiO$_3$ values mentioned in Table 1. In this supplement we have extended the information in the text and addressed some of the key assumptions.

**Appendix A. Effect of termination, rumpling and stoichiometry**

For thin films and nano-scale systems surface termination plays an important role. The (100) stacking in EuTiO$_3$ follows EuO/TiO$_2$/EuO/TiO$_2$/…., similar to other A$^{+2}$B$^{+4}$O$_3$ systems (SrTiO$_3$, BaTiO$_3$, PbTiO$_3$). Behera *et al.* [44] characterized the effect of (100) surface termination on atomic relaxation and polarization in PbTiO$_3$ thin films. For PbO- and TiO$_2$-terminations, the atoms at the surface relax towards the bulk of the thin film. Similar surface relaxation of atoms is expected for nanowires, where the atoms relax toward the center of the wire. Figure S1(a) illustrates the bulk EuTiO$_3$. The dotted circle is an example of a nanowire that can be defined from the bulk systems. Figure S1(b) shows the relaxation of atoms towards the center of the wire. For our calculations, we have assumed that all the distances between the atoms decreases uniformly in the transverse directions (towards the wire radii) in comparison with the bulk (a), while in z-direction (in and out of the paper) the distances increases (in agreement with Eq.(2) and Fig.2 from the manuscript). This assumption seems consistent with continuum theory we used. Figure S1(c) shows the atomic arrangement of a typical EuTiO$_3$ nanowire.

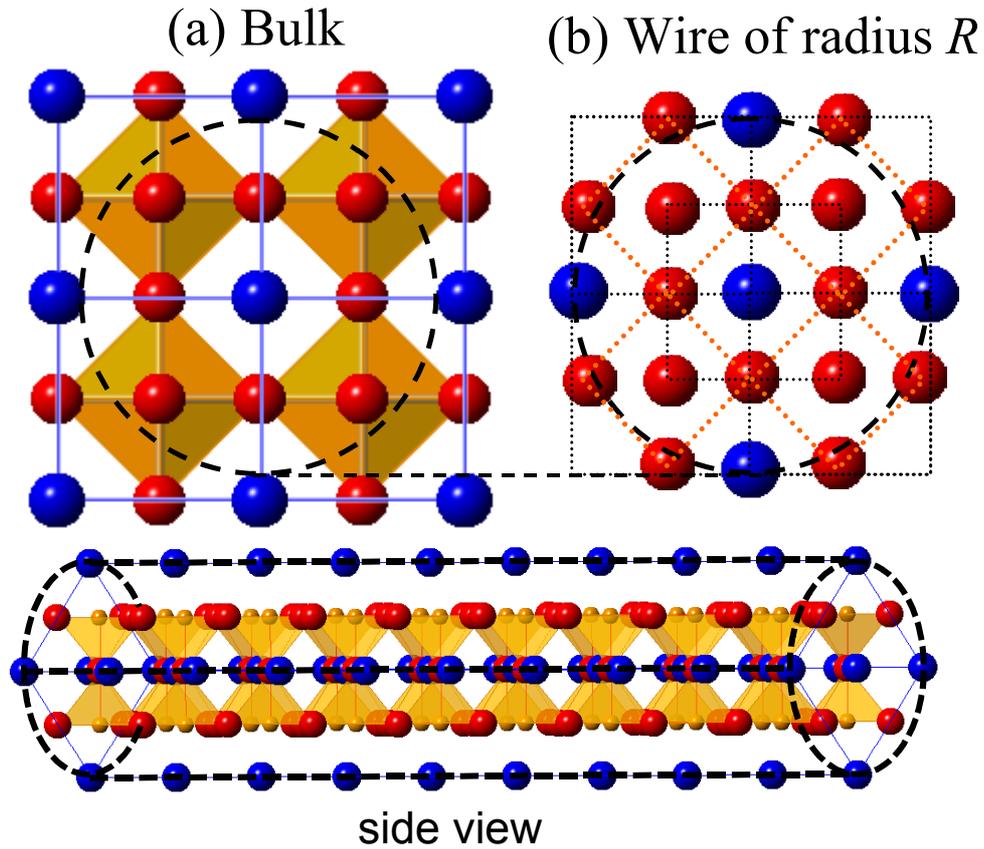

**Fig. S1** Schematic of (a) bulk EuTiO$_3$, where the dotted circle indicates the definition of a nanowire, (b) a nanowire with atoms surface atoms relaxing towards the center of the wire, (c) the side view of a EuO-terminated nanowire.

In addition to surface relaxation, surface rumpling is another important issue for discussion. For PbTiO$_3$ "unpolarized" thin films, surface rumpling develops a polarization of ~5% of the bulk polarization [44]. Similar polarization values (0.03 C/m$^2$) pointing towards the bulk is also reported by Speliarsky *et al.*[45] for four unit cell PbTIO$_3$ thin films. For the polarizations characterized in the manuscript for nanowires (0.1-0.5 C/m$^2$), this ~5% of polarization towards the wire radii induced due to surface rumpling is neglected.

Following the dotted circles in Fig. S1(a) and (b) for bulk EuTiO$_3$, the terminations in EuTiO$_3$ nanowire will follow the (110) stacking, i.e. surfaces are either EuTiO-terminated or OO-terminated. These surfaces are polar in nature and require additional surface reconstruction. Stoichiometry is also another important factor that is to be addressed. However, there is a lack of experimental characterization about these individual contributions for nanowires and not many first-principles calculations on EuTiO$_3$ are available in the literature. Therefore, our current model do not explicitly address these issues, however, includes these effects in the surface stress parameter used for the calculations.

**Appendix B. Possible phases of multiferroic and their stability**

Free energy of isotropic multiferroic in the forth power approximation is

$$F = \frac{\alpha_{PR}(T)}{2}P^2 + \frac{\beta_P}{4}P^4 + \frac{P^2}{2}\left(\eta_{FM}M^2 + \eta_{AFM}L^2\right) + \\ + \frac{\alpha_{MR}(T)}{2}M^2 + \frac{\alpha_{LR}(T)}{2}L^2 + \frac{\beta_M}{4}M^4 + \frac{\beta_L}{4}L^4 + \frac{\lambda}{2}L^2M^2 \quad (B.1)$$

It depends on polarization $P$, magnetization $M$ and "antimagentization" $L$ (order parameter of antiferroelectric phase). This free energy is stable at high values of order parameters only under the conditions:

$$\beta_P > 0, \ \beta_M > 0, \ \beta_L > 0, \ \sqrt{\beta_M \beta_P} + \eta_{FM} > 0, \ \sqrt{\beta_P \beta_L} + \eta_{AFM} > 0, \ \sqrt{\beta_M \beta_L} + \lambda > 0, \quad (B.2)$$

Possible phases are listed below

**1. Para phase**

with $P = M = L = 0$ and free energy $F_{para} = 0$. This phase is stable at $\alpha_{PR} > 0$, $\alpha_{MR} > 0$ and $\alpha_{LR} > 0$.

**2. FE-phase ("P")** with

$P = \pm\sqrt{\dfrac{-\alpha_{PR}}{\beta_P}}$, $M = 0$, $L = 0$ and free energy $F_P = \dfrac{-(\alpha_{PR})^2}{4\beta_P}$. This phase exists and is stable under the conditions $\alpha_{PR} < 0$, $\alpha_{MR} - \eta_{FM}(\alpha_{PR}/\beta_P) > 0$, $\alpha_{LR} - \eta_{AFM}(\alpha_{PR}/\beta_P) > 0$.

**3. FE-FM phase ("PM")** with

$$P = \sqrt{-\frac{\alpha_{PR} - \eta_{FM}(\alpha_{MR}/\beta_M)}{\beta_P - (\eta_{FM}^2/\beta_M)}}, \quad M = \sqrt{-\frac{\alpha_{MR} - \eta_{FM}(\alpha_{PR}/\beta_P)}{\beta_M - (\eta_{FM}^2/\beta_P)}}, \quad L = 0. \quad (B.3)$$

Free energy is $F_{PM} = -\dfrac{\beta_M \alpha_{PR}^2 + \beta_P \alpha_{MR}^2 - 2\eta_{FM}\alpha_{PR}\alpha_{MR}}{4(\beta_P\beta_M - \eta_{FM}^2)}$. This phase exists and is stable under the conditions $-\alpha_{PR} + \eta_{FM}(\alpha_{MR}/\beta_M) > 0$, $-\alpha_{MR} + \eta_{FM}(\alpha_{PR}/\beta_P) > 0$, $\sqrt{\beta_P\beta_M} > \eta_{FM}$ and

$\eta_{AFM}(-\alpha_{PR}\beta_M + \alpha_{MR}\eta_{FM}) + \gamma_{ML}(-\alpha_{MR}\beta_P + \alpha_{PR}\eta_{FM}) + \alpha_{LR}(\beta_P\beta_M - \eta_{FM}^2) > 0$.

**4. FE-AFM phase ("PL")** with

$$P = \sqrt{-\frac{\alpha_{PR} - \eta_{AFM}(\alpha_{LR}/\beta_L)}{\beta_P - (\eta_{AFM}^2/\beta_L)}}, \quad M = 0, \quad L = \sqrt{-\frac{\alpha_{LR} - \eta_{AFM}(\alpha_{PR}/\beta_P)}{\beta_L - (\eta_{AFM}^2/\beta_P)}} \quad (B.4)$$

Free energy is $F_{PL} = -\dfrac{\beta_L \alpha_{PR}^2 + \beta_P \alpha_{LR}^2 - 2\eta_{AFM}\alpha_{PR}\alpha_{LR}}{4(\beta_P\beta_L - \eta_{AFM}^2)}$. This phase exists and is stable under the conditions $-\alpha_{PR} + \eta_{AFM}(\alpha_{LR}/\beta_L) > 0$, $-\alpha_{LR} + \eta_{AFM}(\alpha_{PR}/\beta_P) > 0$, $\beta_P\beta_L - \eta_{AFM}^2 > 0$, $\eta_{FM}(-\alpha_{PR}\beta_L + \alpha_{LR}\eta_{AFM}) + \lambda(-\alpha_{LR}\beta_P + \alpha_{PR}\eta_{AFM}) + \alpha_{MR}(\beta_P\beta_L - \eta_{AFM}^2) > 0$.

**5. FM phase ("M")** with

$$P = 0, \quad M = \sqrt{-\dfrac{\alpha_{MR}}{\beta_M}}, \quad L = 0. \qquad (B.5)$$

Free energy is $F_M = \dfrac{-(\alpha_{MR})^2}{4\beta_M}$. This phase exists and is stable under the conditions $\alpha_{MR} < 0$, $\alpha_{PR} - \eta_{FM}(\alpha_{MR}/\beta_M) > 0$, $\alpha_{LR} - \eta_{AFM}(\alpha_{MR}/\beta_M) > 0$.

**6. AFM phase ("L")** with

$$P = 0, \quad M = 0, \quad L = \sqrt{-\dfrac{\alpha_{LR}}{\beta_L}}. \qquad (B.6)$$

Free energy is $F_L = \dfrac{-(\alpha_{LR})^2}{4\beta_L}$. This phase exists and is stable under the conditions $\alpha_{LR} < 0$, $\alpha_{MR} - \gamma_{ML}(\alpha_{LR}/\beta_L) > 0$, $\alpha_{PR} - \eta_{AFM}(\alpha_{LR}/\beta_L) > 0$.

**7. FM-AFM phase ("ML")** with

$$P = 0, \quad M = \sqrt{-\dfrac{\alpha_{MR} - \lambda(\alpha_{LR}/\beta_L)}{\beta_M - (\lambda^2/\beta_L)}}, \quad L = \sqrt{-\dfrac{\alpha_{LR} - \lambda(\alpha_{MR}/\beta_M)}{\beta_L - (\lambda^2/\beta_M)}}. \qquad (B.7)$$

Free energy is $F_{ML} = -\dfrac{\beta_L \alpha_{MR}^2 + \beta_M \alpha_{LR}^2 - 2\gamma_{ML}\alpha_{MR}\alpha_{LR}}{4(\beta_M\beta_L - \gamma_{ML}^2)}$. This phase exists and is stable under the conditions $-\alpha_{LR} + \lambda(\alpha_{MR}/\beta_M) > 0$, $-\alpha_{MR} + \lambda(\alpha_{LR}/\beta_L) > 0$, $\sqrt{\beta_M\beta_L} > \lambda$ and $\eta_{AFM}(-\alpha_{LR}\beta_M + \alpha_{MR}\lambda) + \eta_{FM}(-\alpha_{MR}\beta_L + \alpha_{LR}\lambda) + \alpha_{PR}(\beta_M\beta_L - \lambda^2) > 0$.

**8. FE-FM-AFM phase ("PML")** with

$$P = \sqrt{-\dfrac{\alpha_{PR}(\beta_M\beta_L - \lambda^2) + \eta_{AFM}(-\alpha_{LR}\beta_M + \alpha_{MR}\lambda) + \eta_{FM}(-\alpha_{MR}\beta_L + \alpha_{LR}\lambda)}{\beta_P\beta_M\beta_L - \beta_P\lambda^2 - \beta_M\eta_{AFM}^2 - \beta_L\eta_{FM}^2 + 2\lambda\eta_{AFM}\eta_{FM}}}, \qquad (B.8a)$$

$$M = \sqrt{-\dfrac{\alpha_{MR}(\beta_P\beta_L - \eta_{AFM}^2) + \eta_{FM}(-\alpha_{PR}\beta_L + \alpha_{LR}\eta_{AFM}) + \gamma_{ML}(-\alpha_{LR}\beta_P + \alpha_{PR}\eta_{AFM})}{\beta_P\beta_M\beta_L - \beta_P\lambda^2 - \beta_M\eta_{AFM}^2 - \beta_L\eta_{FM}^2 + 2\lambda\eta_{AFM}\eta_{FM}}}, \qquad (B.8b)$$

$$L = \sqrt{-\frac{\alpha_{LR}(\beta_P\beta_M - \eta_{FM}^2) + \eta_{AFM}(-\alpha_{PR}\beta_M + \alpha_{MR}\eta_{FM}) + \gamma_{ML}(-\alpha_{MR}\beta_P + \alpha_{PR}\eta_{FM})}{\beta_P\beta_M\beta_L - \beta_P\lambda^2 - \beta_M\eta_{AFM}^2 - \beta_L\eta_{FM}^2 + 2\lambda\eta_{AFM}\eta_{FM}}} . \quad \text{(B.8c)}$$

Free energy is

$$F_{PML} = -\frac{\begin{pmatrix}\alpha_{MR}^2(\beta_M\beta_L - \lambda^2) + \alpha_{MR}^2(\beta_P\beta_L - \eta_{AFM}^2) + \alpha_{LR}^2(\beta_P\beta_M - \eta_{FM}^2) + \\ 2\alpha_{LR}\alpha_{PR}(\eta_{FM}\lambda - \beta_M\eta_{AFM}) + 2\alpha_{MR}\alpha_{PR}(\eta_{AFM}\lambda - \beta_L\eta_{FM}) + 2\alpha_{MR}\alpha_{LR}(\eta_{AFM}\eta_{FM} - \beta_P\lambda)\end{pmatrix}}{4(\beta_P\beta_M\beta_L - \beta_P\lambda^2 - \beta_M\eta_{AFM}^2 - \beta_L\eta_{FM}^2 + 2\lambda\eta_{AFM}\eta_{FM})}$$
(B.9)

This phase exists and is stable under the conditions

$$\alpha_{PR}(\beta_M\beta_L - \lambda^2) + \eta_{AFM}(-\alpha_{LR}\beta_M + \alpha_{MR}\lambda) + \eta_{FM}(-\alpha_{MR}\beta_L + \alpha_{LR}\lambda) < 0, \quad \text{(B.10a)}$$

$$\alpha_{MR}(\beta_P\beta_L - \eta_{AFM}^2) + \eta_{FM}(-\alpha_{PR}\beta_L + \alpha_{LR}\eta_{AFM}) + \lambda(-\alpha_{LR}\beta_P + \alpha_{PR}\eta_{AFM}) < 0, \quad \text{(B.10b)}$$

$$\alpha_{LR}(\beta_P\beta_M - \eta_{FM}^2) + \eta_{AFM}(-\alpha_{PR}\beta_M + \alpha_{MR}\eta_{FM}) + \lambda(-\alpha_{MR}\beta_P + \alpha_{PR}\eta_{FM}) < 0, \quad \text{(B.10b)}$$

$$\beta_P\beta_M\beta_L - \beta_P\lambda^2 - \beta_M\eta_{AFM}^2 - \beta_L\eta_{FM}^2 + 2\lambda\eta_{AFM}\eta_{FM} > 0. \quad \text{(B.10c)}$$

**Appendix C. The case of EuTiO$_3$**

Katsufuji and Takagi [1] reported about temperature and magnetic field dependence of dielectric permittivity ε and magnetic susceptibility of EuTiO$_3$ (ETO) in the range $T = 2\text{-}100$ K and magnetic fields from 0 to 5 Tesla. From these data they deduced temperature dependence of $\alpha_P$ (Barret equation) and $\alpha_M = \alpha_{MT}(T - T_C)$ corresponding to the bulk para-phase.

Katsufuji and Takagi [1] fitted strong changes of ε at the antiferromagnetic phase of ETO below 5.5 K using average spin calculated from the mean-field model. We could use phenomenological model from **Appendix B**, which predicts the following dependence of dielectric permittivity for the AFM phase:

$$\varepsilon = \varepsilon_b + \frac{1}{\varepsilon_0(\alpha_P(T) + \eta_{AFM}L^2)} . \quad \text{(C.1)}$$

Next we should find the values of order parameter of AFM phase as:

$$L = \sqrt{-\frac{\alpha_L}{\beta_L}} . \quad \text{(C.2)}$$

Following Katsufuji and Takagi [1] we could assume that for $T \to 0$ the spins of the sublattices are saturated to 7/2 per unit cell (but have alternating signs), and get value of $L$ (as a half-sum of the sublattices magnetizations) as $L = 0.509 \cdot 10^6$ A/m ($\mu_B (7/2)/(4 \cdot 10^{-10}\text{m})^3$). Using this estimation we could get the coupling constant $\eta_{AFM} = 0.16 \cdot 10^{-3}$.

Assuming that temperature coefficients of $\alpha_M$ and $\alpha_L$ are the same, i.e. $\alpha_M = \alpha_{MT}(T - T_C)$ and $\alpha_L = \alpha_{MT}(T - T_N)$, where $T_C$ and $T_N$ are the Curie and Néel temperatures of ETO respectively, we could estimate $\beta_L$ to be $1.3*10^{-16}$ SI units.

In order to estimate the remaining magnetic parameters, $\beta_M$ and $\gamma_{ML}$, we should recall magnetic field dependences of magnetization $M$ and "antimagnetization" $L$. Using the equations:

$$\alpha_M(T)M + \beta_M M^3 + \lambda L^2 M - \mu_0 H = 0, \qquad \alpha_L(T)L + \beta_L L^3 + \lambda L M^2 = 0. \tag{C.3}$$

and with the condition that at magnetic field about 1 Tesla, $M \approx L$ [1] we estimated $\beta_M = 0.8 \times 10^{-16}$, $\lambda = 1.0 \times 10^{-16}$. Expansion coefficient $\beta_P$, determining spontaneous polarization of ferroelectric (FE) phase, could be estimated from the first principles calculations of Fennie and Rabe [3] and Lee *et al.* [4], who considered epitaxial films of ETO on different substrates. They found maximal spontaneous polarization of ~0.2–0.3 C/m² in compressed films and critical misfit strains for the transition between AFM and FE+FM phases.

This fitting also allows us to estimate electrostriction and magnetostriction constants, since the critical value of misfit strain govern the transitions between paraelectrics-antiferromagnetic and ferroelectric-ferromagnetic phases are calculated from the first principles by Lee *et al.* [4]. Since they found no intermediate phases (like ferroelectric-antiferromagnetic one), theses findings allows us to determine unambiguously all the set of striction constants.

**Appendix D. Elastic constants of EuTiO₃**

For cubic systems the following relation should hold true:

$$B = \frac{(c_{11} + 2c_{12})}{3} \tag{D.1}$$

We extracted the Bulk modulus $B$ from Ranjan et al [37] as $(172.6 \pm 10)$ GPa.

For the sake of simplicity we suppose that we could approximate a solid as elastically isotropic, i.e.

$$c_{11} = \frac{Y(1-\nu)}{(1+\nu)(1-2\nu)}, \quad c_{12} = \frac{Y\nu}{(1+\nu)(1-2\nu)}, \quad c_{44} = \frac{Y}{2(1+\nu)}, \quad \frac{c_{12}}{c_{11}} = \frac{\nu}{1-\nu}, \quad \nu = \frac{c_{12}}{c_{11}+c_{12}} \tag{D.2}$$

where Y is Young module and $\nu$ is the Poisson coefficient. Thus bulk modulus could be rewritten as $B = c_{11}\frac{(1+\nu)}{3(1-\nu)}$.

Assuming the Poisson ratio $\nu$ as 0.24 (equal to that of SrTiO₃ [46] and close to values of BaTiO₃ and PbTiO₃ [47]), we could get stiffness component $c_{11}$ as listed in the **Table 1** of the main text and Young module values, then remaining components, $c_{12}$ and $c_{44}$, could be estimated. Elastic compliances, $s_{ij}$ could be then calculated as the inverse matrix.